\documentstyle[12pt]{article}
\newcommand{\dd}{{\mbox{d}}}
\newcommand{\matr}[1]{\mbox{#1}}
\newcommand{\Li}{\mbox{Li}_2}
\newcommand{\vecc}[1]{\mbox{\boldmath $#1$}}

\title{Radiative corrections to the background \\
       of $\mu\to e\gamma$ decay}

\author{
A.B.~Arbuzov$^{1}$, O.~Krehl$^{2}$, E.A.~Kuraev$^{1}$, \\
E.~Magar$^{1}$, and B.G.~Shaikhatdenov$^{1}$ \\
$^{1}$ {\it Joint Institute for Nuclear Research, Dubna, 141980  Russia} \\
$^{2}$ {\it Institut f\"ur Kernphysik, Forschungzentrum J\"ulich GmbH} \\
{\it 52425 J\"ulich, Germany }
}

\date{}
\begin{document}
\setcounter{page}{-1}
\maketitle

\begin{abstract}
Radiative muon decay in the kinematics similar to the neutrinoless
decay $\mu\to e\gamma$ is considered.
Radiative corrections due to one--loop virtual photons and
emission of additional soft or hard photons are taken into account.
Analytical expressions and numerical estimations are presented.
\end{abstract}

\vspace{1cm}
\thispagestyle{empty}
\begin{center}
{\large \bf Радиационные поправки к фоновому процессу
для распада $\mu\to e\gamma$ }
\end{center}

\vspace{.2cm}
\begin{center}
{ \bf А.Б.~Арбузов$^{1}$, О. Крел$^{2}$, Э.А.~Кураев$^{1}$, Е.Н. Магар$^{1}$,
Б.Г.~Шайхатденов$^{1}$}
\end{center}
\vspace{.3cm}
\begin{center}
{
$^{1}$ \it Объединенный институт ядерных исследований, Дубна, 141980, Россия \\
$^{2}$ {\it Institut f\"ur Kernphysik, Forschungzentrum J\"ulich GmbH} \\
{\it 52425 J\"ulich, Germany } } \\[.2cm]
\end{center}
\vspace{.3cm}

\begin{center}
{\bf Аннотация}
\end{center}
Радиационный распад мюона рассмотрен в кинематике, имитирующей
безнейтринный распад $\mu\to e\gamma$. Вычислены радиационные
однопетлевые поправки с учетом излучения дополнительных
мягкого или жесткого фотонов. Представлены аналитические
выражения и даны численные оценки.

\thispagestyle{empty}
\newpage
\setcounter{page}{1}

\section{Introduction}

Since the discovery of the muon in 1936 its relation to electron is
a puzzle. Really, the only difference between these two elementary
particles is in their masses. The lepton number conservation law
has no deep sources in space--time properties or gauge theories.
Moreover many extensions of the Standard Model predict processes with
violation of this law ($\mu\to e\gamma,\; e\gamma\gamma,\; e\bar{e}e$ etc.).
Intensive search of these extensions was performed in 1977~\cite{r1}.
The modern state of the subject is elucidated in papers~\cite{exp,Kuno}
and references therein.
Indeed, if there is a unification of
quarks and leptons, then the existence of $b\to s\gamma$ decay
leads to that of $\mu\to e\gamma$. Different models give
a wide range of predictions for the branching ratio of this
neutrinoless muon decay.
The present experimental upper limit~\cite{exp} on the branching ratio is
\begin{eqnarray}
B=\frac{\Gamma(\mu\to e\gamma)}{\Gamma_{\mu}^{tot}} < 4.9\cdot 10^{-11}.
\end{eqnarray}
This value imposed already strong restrictions on parameters
of supersymmetric~\cite{gut,ngut} and other models~\cite{moh}.
In the model independent approach~\cite{chi} one gets
boundaries on parameters of possible structures in
the matrix element of the muon decay.
Several new experiments
are planned to improve the precision. They will either find
the decay or put much more stronger restrictions and even
discriminate some models.
The forthcoming experiment at PSI (if doesn't find the decay) will
put the limit on the $\mu\to e\gamma$
decay branching ratio of about $5\cdot 10^{-14}$. Another
experiment is proposed at BNL, where they are going to
reach the level of $10^{-16}$. These experiments are very
important, since they have rather wide possibilities for the
search of new physics comparable with those of high
energy colliders.
In this paper we consider the important background process
\begin{equation}
\mu(p) \to e(p_2)+\gamma(k_1)+(\nu_{\mu}+\bar \nu_e)(q)
\end{equation}
in the kinematical situation, imitating the neutrinoless decay.
Namely, we suppose
\begin{eqnarray*}
n=\frac{2p_2q}{M^2} \sim l=\frac{2k_1q}{M^2} \sim \sqrt{Q^2}
=\sqrt{q^2/M^2}\ll 1,
\end{eqnarray*}
where $q$ is the 4--momentum carried by neutrinos, and $M$ is the muon mass.
The width in the lowest order of perturbation theory
was calculated many years ago~\cite{r5}.
The expression for the width reads:
\begin{eqnarray} \label{Bo}
\dd\Gamma_{\mathrm{Born}}^{\mu \to e \nu \bar \nu \gamma} &=&
\frac{2\alpha G_F^2}{6(2\pi)^6 M}
\frac{\dd^3 p_2 \dd^3 k_1}{\varepsilon_2\omega_1}
\biggl[-\left(\frac{M^4}{2}-q^2\left(q^2-\frac{M^2}{2}
\right)\right)
\biggl(\frac{p}{pk_1}-\frac{p_2}{p_2k_1}\biggr)^2 \nonumber \\
&+& 4q^2 + \frac{(k_1q)^2}{(p_2k_1)(pk_1)}
(2q^2+M^2)\biggr]\, , \quad
q=p-p_2-k_1, \quad \omega_1 = k_1^0.
\end{eqnarray}
The validity of this formula may be confirmed in the limiting case of
soft photon. A multiplier $2$ was lost on right hand side (rhs) of expression
for the width in~\cite{r5}.
The polarized muon radiative decay was considered in~\cite{FU}, and
as a background to the neutrinoless decay it was extensively
discussed in Ref.~\cite{Kuno}.

\section{Radiative corrections}

In the {\it imitating kinematics} (IK) we introduce the relative
energy deviations of the hard electron and photon from $M/2$ and the
acollinearity angle $\theta$:
\begin{equation} \label{sigma}
\sigma_1=1-\frac{2\omega_1}{M}, \qquad \sigma_2=1-\frac{2\varepsilon_2}{M},
\qquad \theta=\widehat{\vecc{p}_2,-\vecc{k}}_1.
\end{equation}
Here we suggest
\begin{equation}
\sigma_1 \sim \sigma_2 \sim \theta \ll 1.
\end{equation}
Rearranging the phase volume
\begin{eqnarray*}
\dd\Phi = \frac{\dd^3p_2 \dd^3k_1}{\omega_1\varepsilon_2}
= 8\pi^2\left(\frac{M}{2}\right)^4(1-\sigma_1)(1-\sigma_2)
\dd\sigma_1\dd\sigma_2 \theta\dd\theta,
\end{eqnarray*}
and expanding the expression for the width in the Born approximation~\cite{FU},
we obtain:
\begin{eqnarray} \label{Bo1}
\frac{\dd\Gamma_{\mathrm{Born}}}{\dd\sigma_1 \dd\sigma_2 \theta\dd\theta} &=&
\frac{\dd\Gamma_0}{\dd\sigma_1 \dd\sigma_2 \theta\dd\theta}
(1 + \delta_1), \\ \nonumber
\frac{\dd\Gamma_0}{\dd\sigma_1 \dd\sigma_2 \theta\dd\theta} &=&
\frac{\alpha G_F^2 M^5}{3\cdot 2^7\pi^4}R, \qquad
R = \sigma^2_2(1+\xi)+\left(4\sigma_1\sigma_2-\frac{\theta^2}{2}\right)(1-\xi)
-\sigma_2\theta \eta, \\ \nonumber
\xi&=&s\cos(\widehat{\vecc s,\vecc p}_2),
\quad \eta=s\sin(\widehat{\vecc s,\vecc p}_2)\cos\varphi, \quad
\vecc{k}_1\vecc{s} = \omega_1(-\xi\cos\theta - \eta\sin\theta), \\ \nonumber
\delta_1&=&\frac{1}{R}\biggl[(-5+3\xi)\sigma_1^2\sigma_2-4(1-\xi)\sigma_2^2
\sigma_1+2(1-\xi)\sigma_1\theta^2+\frac{1}{2}(3-\xi)\sigma_2\theta^2 \\ \nonumber
&+& 4\eta\sigma_1\sigma_2\theta-\frac{5}{4}\eta\theta^3\biggr].
\end{eqnarray}
Here $s$ denotes the spin of the muon, and $\varphi$ is the azimuthal angle
between planes formed by $(\vecc{s},\vecc{p}_2)$ and
$(\vecc{s},\vecc{k}_1)$ in the rest reference frame of the muon.
Note that averaging the above expression over the angle $\varphi$ leads
immediately to the result presented in~\cite{Kuno}.
We shall name higher than second order contributions on the rhs of (\ref{Bo})
(and $\delta_1$ in rhs of (\ref{Bo1})) as {\it relativistic} corrections.
In this paper we will consider the radiative corrections to this width bearing
in mind virtual corrections described by the Feynman diagrams drawn in Fig.~1
together with those arising from the emission of additional soft and hard
photons.

For the measurement of an additional hard photon emission, two
cases have to be considered: with and without external magnetic filed.
In the case without magnetic field
the additional hard photon, moving along the final electron trajectory
within a small angle, which is equal to the detector angular resolution,
is registered together with the electron. In
the opposite case (with magnetic field) those events will be rejected from
statistics due to criterion:
the energy of the electron is less than the maximum energy within some
accuracy.
The standard calculation of one--loop virtual corrections can be considerably
simplified by using the IK features. Some details of our calculations (traces,
vertices and the Tables of relevant integrals) are given in
Appendices.

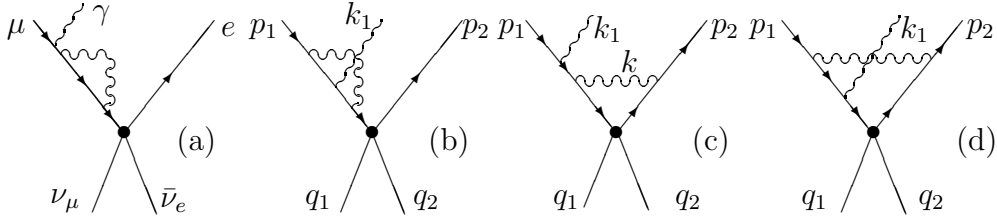
\begin{figure}
\unitlength=0.60mm
\special{em:linewidth 0.4pt}
\linethickness{0.4pt}
\begin{picture}(223.81,64.33)
\put(123.14,59.00){\line(4,-5){20.00}}
\put(143.14,34.00){\line(4,5){20.00}}
\put(135.14,45.51){\oval(2.00,2.33)[t]}
\put(137.14,45.67){\oval(2.00,2.67)[b]}
\put(139.14,45.51){\oval(2.00,2.33)[t]}
\put(141.14,45.67){\oval(2.00,2.67)[b]}
\put(143.14,45.51){\oval(2.00,2.33)[t]}
\put(145.14,45.67){\oval(2.00,2.67)[b]}
\put(147.14,45.51){\oval(2.00,2.33)[t]}
\put(149.14,45.67){\oval(2.00,2.67)[b]}
\put(151.14,45.51){\oval(2.00,2.33)[t]}
\put(125.80,55.67){\vector(3,-4){1.67}}
\put(138.80,39.67){\vector(3,-4){1.67}}
\put(179.81,59.00){\line(4,-5){20.00}}
\put(199.81,34.00){\line(4,5){20.00}}
\put(187.81,50.51){\oval(2.00,2.33)[t]}
\put(189.81,50.67){\oval(2.00,2.67)[b]}
\put(191.81,50.51){\oval(2.00,2.33)[t]}
\put(193.81,50.67){\oval(2.00,2.67)[b]}
\put(195.81,50.51){\oval(2.00,2.33)[t]}
\put(197.81,50.67){\oval(2.00,2.67)[b]}
\put(199.81,50.51){\oval(2.00,2.33)[t]}
\put(201.81,50.67){\oval(2.00,2.67)[b]}
\put(203.81,50.51){\oval(2.00,2.33)[t]}
\put(205.81,50.67){\oval(2.00,2.67)[b]}
\put(207.81,50.51){\oval(2.00,2.33)[t]}
\put(209.81,50.67){\oval(2.00,2.67)[b]}
\put(211.81,50.51){\oval(2.00,2.33)[t]}
\put(194.81,41.66){\oval(2.00,3.33)[lt]}
\put(194.81,45.00){\oval(2.00,3.33)[rb]}
\put(196.81,44.66){\oval(2.00,3.33)[lt]}
\put(196.81,48.00){\oval(2.00,3.33)[rb]}
\put(198.81,47.66){\oval(2.00,3.33)[lt]}
\put(198.81,51.00){\oval(2.00,3.33)[rb]}
\put(200.81,50.66){\oval(2.00,3.33)[lt]}
\put(200.81,54.00){\oval(2.00,3.33)[rb]}
\put(202.81,53.66){\oval(2.00,3.33)[lt]}
\put(202.81,57.00){\oval(2.00,3.33)[rb]}
\put(204.81,56.66){\oval(2.00,3.33)[lt]}
\put(204.81,60.00){\oval(2.00,3.33)[rb]}
\
\put(182.47,55.67){\vector(3,-4){1.67}}
\put(189.14,47.67){\vector(3,-4){1.67}}
\put(195.47,39.67){\vector(3,-4){1.67}}
\put(131.80,49.33){\oval(2.00,3.33)[lt]}
\put(131.80,52.67){\oval(2.00,3.33)[rb]}
\put(133.80,52.33){\oval(2.00,3.33)[lt]}
\put(133.80,55.67){\oval(2.00,3.33)[rb]}
\put(135.80,55.33){\oval(2.00,3.33)[lt]}
\put(135.80,58.67){\oval(2.00,3.33)[rb]}
\put(137.80,58.33){\oval(2.00,3.33)[lt]}
\put(130.80,49.67){\vector(3,-4){1.67}}
\put(157.47,51.67){\vector(3,4){1.67}}
\put(146.14,37.67){\vector(3,4){1.67}}
\put(214.14,51.67){\vector(3,4){1.67}}
\put(202.47,37.34){\vector(3,4){1.67}}
\put(120.14,57.00){\makebox(0,0)[cc]{$p_1$}}
\put(175.81,57.00){\makebox(0,0)[cc]{$p_1$}}
\put(167.14,57.00){\makebox(0,0)[cc]{$p_2$}}
\put(223.81,57.00){\makebox(0,0)[cc]{$p_2$}}
\put(146.00,50.00){\makebox(0,0)[cc]{$k$}}
\put(141.14,57.00){\makebox(0,0)[cc]{$k_1$}}
\put(209.14,57.00){\makebox(0,0)[cc]{$k_1$}}
\
\put(14.00,59.00){\line(4,-5){20.00}}
\put(34.00,34.00){\line(4,5){20.00}}
\put(21.00,51.51){\oval(2.00,2.00)[t]}
\put(23.00,51.51){\oval(2.00,2.00)[b]}
\put(25.00,51.51){\oval(2.00,2.00)[t]}
\put(27.00,51.51){\oval(2.00,2.00)[b]}
\put(29.00,51.51){\oval(2.00,2.00)[t]}
\put(31.00,51.51){\oval(2.00,2.00)[lb]}
\put(31.00,49.51){\oval(2.00,2.00)[r]}
\put(31.00,47.51){\oval(2.00,2.00)[l]}
\put(31.00,45.51){\oval(2.00,2.00)[r]}
\put(31.00,43.51){\oval(2.00,2.00)[l]}
\put(31.00,41.51){\oval(2.00,2.00)[r]}
\put(31.00,39.51){\oval(3.00,2.00)[lt]}
\put(16.00,56.50){\vector(3,-4){1.67}}
\put(24.00,46.50){\vector(3,-4){1.67}}
\put(30.00,39.00){\vector(3,-4){1.67}}
\put(43.00,45.25){\vector(3,4){1.67}}
\put(19.80,53.33){\oval(2.00,3.33)[lt]}
\put(19.80,56.67){\oval(2.00,3.33)[rb]}
\put(21.80,56.33){\oval(2.00,3.33)[lt]}
\put(21.80,59.67){\oval(2.00,3.33)[rb]}
\put(23.80,59.33){\oval(2.00,3.33)[lt]}
\put(23.80,62.67){\oval(2.00,3.33)[rb]}
\put(29.00,60.00){\makebox(0,0)[cc]{$\gamma$}}
\put(10.14,57.00){\makebox(0,0)[cc]{$\mu$}}
\put(57.14,57.00){\makebox(0,0)[cc]{$e$}}
\
\put(69.00,59.00){\line(4,-5){20.00}}
\put(89.00,34.00){\line(4,5){20.00}}
\put(76.00,51.51){\oval(2.00,2.00)[t]}
\put(78.00,51.51){\oval(2.00,2.00)[b]}
\put(80.00,51.51){\oval(2.00,2.00)[t]}
\put(82.00,51.51){\oval(2.00,2.00)[b]}
\put(84.00,51.51){\oval(2.00,2.00)[t]}
\put(86.00,51.51){\oval(2.00,2.00)[lb]}
\put(86.00,49.51){\oval(2.00,2.00)[r]}
\put(86.00,47.51){\oval(2.00,2.00)[l]}
\put(86.00,45.51){\oval(2.00,2.00)[r]}
\put(86.00,43.51){\oval(2.00,2.00)[l]}
\put(86.00,41.51){\oval(2.00,2.00)[r]}
\put(86.00,39.51){\oval(3.00,2.00)[lt]}
\put(71.00,56.50){\vector(3,-4){1.67}}
\put(85.80,38.33){\vector(3,-4){1.67}}
\put(98.00,45.25){\vector(3,4){1.67}}
\put(82.00,44.00){\oval(2.00,3.00)[lt]}
\put(82.00,47.00){\oval(2.00,3.00)[rb]}
\put(84.00,47.00){\oval(2.00,3.00)[lt]}
\put(84.00,50.00){\oval(2.00,3.00)[rb]}
\put(86.00,50.00){\oval(2.00,3.00)[lt]}
\put(86.00,53.00){\oval(2.00,3.00)[rb]}
\put(88.00,53.00){\oval(2.00,3.00)[lt]}
\put(88.00,56.00){\oval(2.00,3.00)[rb]}
\put(90.00,56.00){\oval(2.00,3.00)[lt]}
\put(90.00,59.00){\oval(2.00,3.00)[rb]}
\put(86.00,60.00){\makebox(0,0)[cc]{$k_1$}}
\put(65.00,57.00){\makebox(0,0)[cc]{$p_1$}}
\put(112.00,57.00){\makebox(0,0)[cc]{$p_2$}}
\put(89.00,34.00){\line(-2,-5){7.00}}
\put(89.00,34.00){\line(2,-5){7.22}}
\put(21.00,19.00){\makebox(0,0)[cc]{$\nu_{\mu}$}}
\put(45.00,19.00){\makebox(0,0)[cc]{$\bar\nu_{e}$}}
\put(77.00,19.00){\makebox(0,0)[cc]{$q_1$}}
\put(101.00,19.00){\makebox(0,0)[cc]{$q_2$}}
\put(132.00,19.00){\makebox(0,0)[cc]{$q_1$}}
\put(34.00,34.00){\line(-2,-5){7.00}}
\put(143.00,35.00){\line(-2,-5){7.00}}
\put(200.00,34.00){\line(-2,-5){7.00}}
\put(34.00,34.00){\line(2,-5){7.22}}
\put(143.00,35.00){\line(2,-5){7.22}}
\put(200.00,34.00){\line(2,-5){7.22}}
\put(159.00,19.00){\makebox(0,0)[cc]{$q_2$}}
\put(187.00,19.00){\makebox(0,0)[cc]{$q_1$}}
\put(213.00,19.00){\makebox(0,0)[cc]{$q_2$}}
\put(0.00,0.00){\line(0,0){0.00}}
\put(0.00,0.00){\line(0,0){0.00}}
\put(34.00,34.00){\line(-4,5){20.00}}
\put(13.00,58.00){\line(0,0){0.00}}
\put(34.00,34.00){\makebox(0,0)[cc]{$\bullet$}}
\put(89.00,34.00){\makebox(0,0)[cc]{$\bullet$}}
\put(143.00,34.00){\makebox(0,0)[cc]{$\bullet$}}
\put(200.00,34.00){\makebox(0,0)[cc]{$\bullet$}}
\put(50.00,32.00){\makebox(0,0)[cc]{(a)}}
\put(106.00,32.00){\makebox(0,0)[cc]{(b)}}
\put(164.00,32.00){\makebox(0,0)[cc]{(c)}}
\put(223.00,32.00){\makebox(0,0)[cc]{(d)}}
\end{picture}
\caption{The subset of Feynman diagrams for radiative muon decay.}
\end{figure}

Ultraviolet divergences of loop integrals are eliminated in a standard way
using the
renormalization constants of the wave functions of electron and muon:
\begin{eqnarray}
Z_{1e} &=& 1-\frac{\alpha}{2\pi}\biggl[\frac{1}{2}\ln\frac{\Lambda^2}{m^2}+
\ln\frac{\lambda^2}{m^2}+\frac{9}{4}\biggr],\\ \nonumber
Z_{1\mu} &=& 1-\frac{\alpha}{2\pi}\biggl[\frac{1}{2}\ln\frac{\Lambda^2}{M^2}+
\ln\frac{\lambda^2}{M^2}+\frac{9}{4}\biggr],
\end{eqnarray}
where $m,\lambda,\Lambda$ are the electron mass, infrared and ultraviolet
cut--off momentum parameters, $(\lambda \ll m\,,\Lambda \gg M)$.
The final result for the one--loop virtual corrections reads:
\begin{eqnarray}
\frac{\dd\Gamma^{\mathrm{virt}}}{\dd\sigma_1 \dd\sigma_2 \theta\dd\theta}
&=&\frac{\dd\Gamma_0}{\dd\sigma_1 \dd\sigma_2 \theta\dd\theta}
\delta_V , \\ \nonumber
\delta_V &=& \frac{\alpha}{\pi R}
\biggl\{ R \biggl( \frac{3}{2}L - (L-1)\ln\frac{Mm}{\lambda^2}
+ \frac{\pi^2}{6} \biggr)
+ \frac{\sigma_1^2}{4}(1-\xi) \\ \nonumber
&+& \sigma_2^2 \biggl( - 3 - 2L + \frac{\pi^2}{6}\biggr)(1+\xi)
+ \sigma_1\sigma_2\biggl[-\frac{23}{4} + \xi\biggl( \frac{9}{4}
+ 4L + \frac{2\pi^2}{3} \biggr)\biggr] \\ \nonumber
&+& \theta^2\biggl[\frac{13}{16} - \xi\biggl( \frac{3}{16}
+ \frac{1}{2}L + \frac{\pi^2}{12} \biggr)\biggr]
+ \sigma_2\theta \eta\biggl(\frac{7}{4}+2L\biggr)
- \frac{\pi^2}{12}\sigma_1\theta \eta \biggr\},
  \\ \nonumber
L&=&\ln\frac{M}{m}.
\end{eqnarray}
Taking into account the emission of additional soft photon requires
some care. The reason is that the energy--momentum carried by soft photons
as well as by neutrinos cannot in principle be distinguished in
the experiment. We introduce some small energy fraction parameter
$\Delta_1=2\omega_{\mathrm{soft}}/M  \ll \sigma_1,\sigma_2,\theta$ which
should not affect on observable quantities and actually cancels out
in the final result.
Emission of an additional soft photon, having energy lesser than
$M\Delta_1/2$, can be taken into account in
a usual way~\cite{r6}. The corresponding expression looks as follows:
\begin{equation}
\frac{\dd\Gamma^{\mathrm{soft}}}{\dd^3p_2\dd^3k_1}=\frac{\dd\Gamma_0}
{\dd^3p_2\dd^3k_1}\delta_S, \quad \delta_S = \frac{\alpha}{\pi}
\biggl[2(L-1)\ln\frac{2\omega_{\mathrm{soft}}}{\lambda}-L^2+L+1
-\frac{\pi^2}{6}\biggr].
\end{equation}
Let us suppose, that a photon with momentum $k_2$, having
energy more than $\omega_{\mathrm{soft}}$, is emitted in such a way
that we have still allowed values of the final electron and hard photon
momenta. In this case the additional photon cannot be called
soft, because it changes the kinematics of the process. We
have to consider the corresponding contribution applying
complete set of kinematical restrictions. The main
condition is that the missing momentum squared must be positive:
\begin{eqnarray}
\tilde{q}^2 = ( p - p_2 - k_1 )^2 > 0,\qquad \tilde{q} = q + k_2.
\end{eqnarray}
Having in mind that the matrix element squared is proportional
to the second power of small neutrino momenta, we can write down
the contribution under consideration in the factorized form
\begin{eqnarray}
\frac{\dd\Gamma^{\gamma}}{\dd^3p_2\dd^3k_1} &=&
\frac{\dd\Gamma_0}{\dd^3p_2\dd^3k_1}\;\frac{1}{R}
\biggl(-\frac{\alpha}{4\pi^2}\biggr)
\int\limits_{\omega_2>\omega_{\mathrm{soft}}}\frac{\dd^3k_2}{\omega_2}
\biggl(\frac{p}{pk_2}-\frac{p_2}{p_2k_2}\biggr)^2\tilde{R}
\Theta(\tilde{q}^2), \\ \nonumber
\tilde{R} &=& 2\tilde{Q}^2 + 2\tilde{l}\tilde{n} + \tilde{l}^2
+ \xi(-2\tilde{Q}^2-2\tilde{l}\tilde{n}+\tilde{l}^2), \\ \nonumber
\tilde{Q}^2 &=& \sigma_1\sigma_2 - \frac{1}{4}\theta^2 - x\sigma_1, \qquad
\tilde{l} = \sigma_2 - x, \qquad \tilde{n} = \sigma_1,\\ \nonumber
\tilde{Q}^2 &=& \frac{2}{M}\tilde{q}^2, \qquad
\tilde{l} = \frac{2\tilde{q}p_2}{M}, \qquad
\tilde{n} = \frac{2\tilde{q}k_1}{M},\qquad
x = \frac{2\omega_2}{M}, \quad \omega_2 = k_2^0.
\end{eqnarray}
The difference in respect to the case of pure soft photon
emission is that we have the {\it shifted} quantity $\tilde{R}$
instead of the Born one ($R$) under the integral sign.
The above expression guarantees that the energies and angles
of the observed electron and photons are the same as defined
in (\ref{sigma}).
Transforming the above formula we get
\begin{eqnarray}
\delta_\gamma&=&\frac{\alpha}{2\pi} \Biggl\{
\int\limits_{\Delta_1}^{x_{\mathrm{max}}}
\frac{\dd x}{x}
\tilde{R}(x,c_2=1)\left[-2+4\ln\left(\frac{M\theta_0}{2m}\right)\right]
\Theta\left(\sigma_1\sigma_2-\frac{\theta^2}{4}-x\sigma_1\right)
\nonumber \\ \nonumber
&+& \int\limits_{0}^{2\pi}\frac{\dd\varphi_2}{2\pi}
\int\limits_{-1}^{1-\theta_0^2/2}
\dd c_2\int\limits_{\Delta_1}^{x_{\mathrm{max}}}\frac{\dd x}{x}\;
\tilde{R}(x,c_2,\varphi_2)
\left(-1+\frac{2}{1-c_2}\right)\Theta\biggl(\sigma_1\sigma_2
-\frac{\theta^2}{4} \\ \label{delg}
&-& \frac{x}{2}(\sigma_1+\sigma_2+c_2(\sigma_1-\sigma_2)
-\theta\sqrt{1-c_2^2}\cos\varphi_2) \biggr)\Biggr\}, \\ \nonumber
c_2 &=& \cos(\widehat{\vecc{k}_2,\vecc{p}}_2), \qquad
x_{\mathrm{max}} = \frac{1}{2}\biggl( \sigma_1 + \sigma_2
+ \sqrt{(\sigma_1-\sigma_2)^2+\theta^2}\biggr).
\end{eqnarray}
In this expression we introduced an auxiliary parameter
$\theta_0$ in order to separate the contribution, when the
additional photon is emitted collinear to the electron
momentum; $\theta_0 \ll 1$. So, the first term of Eq.~(\ref{delg})
can be integrated analytically in order to keep track of the
leading logarithmic part. We checked that the final expression
does not depend on $\theta_0$.

Then we arrive to the total answer, that has the form
\begin{equation} \label{fac}
\frac{\dd\Gamma}{\dd\Gamma_0}=1+\delta_1 + \delta_V
+ \delta_S + \delta_{\gamma}\, .
\end{equation}
The dependence on the soft photon parameter
$\Delta_1$ cancels out in the sum $\delta_S + \delta_{\gamma}$
whereas the fictitious photon mass $\lambda$ disappears in the sum
$\delta_S + \delta_V$.

If the experimental set--up does not distinguish in the
detector an electron with a collinear photon, we have to
modify our results in the following way. Let $\theta_0$
define the aperture of the narrow cone, within which the
two particles would be detected as a unique one. Then we
should take the {\it non--shifted} value for $R$ in the
first integral of Eq.~(\ref{delg}). We have to add also
the rest contribution of hard photon emission within the same
cone. It can be obtained using the quasireal
electron method~\cite{r7}:
\begin{eqnarray}
\frac{\dd\Gamma^{\mathrm{hard}}}{\dd\sigma_1 \dd\sigma_2 \theta\dd\theta} &=&
\frac{\dd\Gamma_0}{\dd\sigma_1 \dd\sigma_2 \theta\dd\theta}\,
\frac{\alpha}{\pi} \int\limits_{x_{\mathrm{max}}'}^{1}\dd x\,
\frac{1+(1-x)^2}{x}\ln\left(\frac{M\theta_0}{2m}\right), \\ \nonumber
x_{\mathrm{max}}' &=& \sigma_2 - \frac{\theta^2}{4\sigma_1}\, .
\end{eqnarray}
The lower limit comes here from the $\Theta$-function
in the first integral of Eq.~(\ref{delg}).

In the presence of a magnetic field, when the electron
trajectory is curve, the above expression will give a
part of the background to
the process $\mu \to e \gamma \gamma $, considered in paper~\cite{r8}.
Really, the final electron will have the
energy $M(1-x)/2$, whereas the quantities $\sigma_1\,,\sigma_2\,,
\theta$ are the same as in the case of single photon emission.

\section{Conclusions}

In Table~1 we give numerical values for $\delta_1$,
$\delta_{SV\gamma}=\delta_{S}+\delta_{V}+\delta_{\gamma}$
versus $\sigma_1,\sigma_2,\theta$, which characterize the
missing energy and momentum (see Eq.~(\ref{sigma})),
and the degree of muon polarization $\xi$.
For typical expected values of $\sigma_1\sim \sigma_2 \sim
\theta \sim 10^{-2}$ (we give 6 points)
one can see, that the relativistic and QED corrections
should be taken into account on the same footing.
The resulting correction to the Born--level decay width
$\dd\Gamma_0$ (see Eq.~(\ref{Bo1})) is given by the sum
$\delta_{SV\gamma} + \delta_1$.

A measurement of the radiative muon decay in the kinematics
close to that of neutrinoless decay is required to
get an independent normalization for the search of the decay
$\mu\to e\gamma$. For this aim our results, we hope, would be
important.

\begin{table}
\begin{center}
\begin{tabular}{|c|c|c|c|r|r|r|r|r|r|}
\hline
$10^2\sigma_1$&$10^2\sigma_2$&$10^2\theta$
& \multicolumn{3}{c|}{$10^2\delta_1$}
& \multicolumn{3}{c|}{$10^2\delta_{SV\gamma}$}
\\ \hline
& & & $ \xi=0 $ & $ \xi=0.5 $ & $ \xi=-0.5 $
& $ \xi=0 $ & $ \xi=0.5 $ & $ \xi=-0.5 $
\\
\hline
$1.0$&$1.0$&$1.0$&$-1.2$ &$-1.0$&$ -1.3$&$-10.5$&$-10.7$&$-10.3$\\ \hline
$3.0$&$3.0$&$3.0$&$-3.7$ &$-3.0$&$ -4.0$&$-8.3$&$-8.5$&$-8.1$ \\ \hline
$5.0$&$5.0$&$5.0$&$-6.1$ &$-5.0$&$ -6.7$&$-7.2$&$-7.5$&$-7.1$ \\ \hline
$6.0$&$6.0$&$3.0$&$-10.0$&$-8.6$&$-10.8$&$-6.6$&$-6.9$&$-6.5$ \\ \hline
$3.0$&$3.0$&$5.9$&$4.4$  &$ 3.8$&$  4.9$&$-13.1$&$-13.2$&$-13.0$\\ \hline
$4.0$&$4.0$&$3.0$&$-6.0$ &$-5.0$&$ -6.5$&$-7.5$&$-7.8$&$-7.4$ \\ \hline
\end{tabular}
\end{center}
\label{num}
\caption{ Numerical estimations for the corrections $\delta_1$ and
          $\delta_{SV\gamma}$ versus $\sigma_1,\sigma_2,\theta,\xi$}
\end{table}

We would like to mention here result obtained in~\cite{r9} on the background
to the three lepton neutrinoless decay
$\mu^+ \to e^+ e^+ e^-$.
For an experimental set--up when the electron and positron
energies $\varepsilon^{\pm}$ are measured, it reads
\begin{equation}
\frac{\dd\Gamma}{\Gamma_0 \dd \Delta}
= \frac{\alpha^2}{\pi^2}\frac{13}{36}(2-w)^2
\ln\frac{M^2}{m^2},\quad w=\frac{2}{M}(\varepsilon^+_1+\varepsilon^+_2
+\varepsilon^-), \quad w \to 2.
\end{equation}

We have to discuss some features of the results presented. At first we note,
that the large logarithm $L$ does not factorize before the
Born--like structure ($R$), as one may expect. We claim that the
factorization theorem, which was proved for high energy processes,
should not work here. Another problem is
that if one integrated out over the whole phase volume of the
second photon, he would still have in the answer the
logarithm of the mass ratio. Formally, this violates the
Kinoshita--Lee--Nauenberg theorem~\cite{kln}; the formula
is infinite in the limit $m\to 0$.
But again, the conditions of the theorem allow us to say, that
the process of radiative muon decay is a legal exception.
One can see the same situation in radiative muon decay at
the Born level~\cite{r6,ber}.

\subsection*{Acknowledgments}
One of us (E.A.K.) is grateful for hospitality to Institut f\"ur Kernphysik,
J\"ulich, where part of this paper was done and to INTAS 93-239 ext.
for financial support.

\section*{Appendix A. Tables of integrals}
\setcounter{equation}{0}
\renewcommand{\theequation}{A.\arabic{equation}}

Here we put the tables of relevant integrals appearing in the loop momentum
integration. The denominators of amplitudes, which correspond to Feynman
diagrams drawn in Fig.1, have the following form:
\begin{equation}
\begin{array}{c}
(1)=(P-k)^2-M^2,\qquad (2)=(p_2-k)^2-m^2\,,\\
(\bar1)=(p-k_1-k)^2-M^2\approx k^2-2kp_2-M^2\,,\\
(\bar2)=(p_2+k_1-k)^2-m^2\approx k^2-2kp+M^2\,,\\
(0)=k^2-\lambda^2.
\end{array}
\end{equation}
We use a symbol $\approx$ to underline the peculiarity of imitating
kinematics. Namely, working out traces we use
\begin{equation}
p_2^2=k_1^2=0,\quad 2p_2k_1=M^2=1,\quad q=0.
\end{equation}
The scalar integrals considered have a form
\begin{equation}
\int \frac{\dd k}{(i)(j)},\quad \int \frac{\dd k}{(i)(j)(k)},
\quad \int \frac{\dd k}{(i)(j)(k)(l)},\quad \dd k=\frac{\dd^4 k}{i\pi^2}.
\end{equation}
Vector and tensor integrals are parametrized as follows:
\begin{eqnarray}
\int \frac{k^{\mu} \dd k}{N}=c k_1^{\mu}+d p_2^{\mu}\,,\nonumber \\
\int \frac{k^{\mu}k^{\nu} \dd k}{N}=g g^{\mu\nu} + \alpha k_1^{\mu}k_1^{\nu}+
\beta p_2^{\mu}p_2^{\nu}+\gamma (k_1,p_2)^{\mu\nu}\,, \nonumber \\
\int \frac{k^{\mu}k^{\nu}k^{\sigma}\dd k}{N}=(G^{(1)}(g,k_1)+G^{(2)}(g,p_2)+\kappa
(k_1)^3+\tau (p_2)^3+\nonumber \\
\psi(p_2,k_1,k_1)+ \rho(p_2,p_2,k_1))^{\mu\nu\sigma}\,,
\end{eqnarray}
where we denote different symmetrical combinations, for instance:
\begin{equation}
(g,a)^{ijk}=g^{ij}a^k+g^{ik}a^j+g^{jk}a^i\,,(a,b)^{ij}=a^ib^j+a^jb^i,...
\end{equation}
Below we put the values of the coefficients and the scalar integrals.
In the tables $2\div 7$ we used
$Y=\ln\frac{\Lambda^2}{M^2}$,
$L=\ln\frac{M}{m}$, $X=\frac{\pi^2}{6}$, $Z=\ln\frac{Mm}
{\lambda^2}$. All the integrals we put in dimensionless form by
setting $M=1$.

\renewcommand{\baselinestretch}{1.3}

\begin{table}[H]
\begin{center}
\begin{tabular}{||c|c|c|c|c|c|c|c|c||}
\hline
$(01)$&$(0\bar{1})$&$(02)$&$(0\bar{2})$&$(12)$
&$(1\bar{2})$&$(2\bar{1})$&$(2\bar{2})$&$(1\bar{1})$ \\ \hline
$Y+1$&$Y$&$Y+2L+1$&$Y+1$&$Y$&$Y$&$Y$&$Y+2L-1$&$Y-1$ \\
\hline
\end{tabular}
\end{center}
\label{scalar}
\caption{\small Scalar integrals with 2 denominators}
\end{table}

\begin{table}[H]
\begin{center}
\begin{tabular}{||c|c|c|c|c||}
\hline
$(012)$&$(01\bar{1})$&$(0\bar{1}2)$&$(1\bar{1}2)$&$(01\bar{2})$ \\ \hline
$-LZ$&$-X$&$-2L-1$&$-1$&$0$\\
\hline
\end{tabular}
\vspace{0.3cm}\\
\begin{tabular}{||c|c||c|c||}
\hline
$(02\bar{2})$&$(12\bar{2})$&$(01\bar{1}2)$&$(012\bar{2})$ \\ \hline
$2L^2-X$&$1-2L$&$ZL-X$&$-ZL-2L^2+X$ \\
\hline
\end{tabular}
\end{center}
\label{scalar2}
\caption{\small Scalar integrals with 3 and 4 denominators}
\end{table}

\begin{table}[H]
\begin{center}
\begin{tabular}{||c|c|c|c|c|c||}
\hline
&$(01)$&$(0\bar{1})$&$(02)$&$(0\bar{2})$&$(12)$\\ \hline
$d$&$\frac{1}{2}Y-\frac{1}{4}$&$\frac{1}{2}Y-\frac{1}{2}$
&$\frac{1}{2}Y+L-\frac{1}{4}$&$\frac{1}{2}Y+\frac{1}{4}$
&$Y-\frac{1}{2}$\\ \hline
$c$&$\frac{1}{2}Y-\frac{1}{4}$&$0$&$0$
&$\frac{1}{2}Y+\frac{1}{4}$&$\frac{1}{2}Y-\frac{1}{2}$\\
\hline
\end{tabular}
\vspace{0.3cm}\\
\begin{tabular}{||c|c|c|c|c||}
\hline
&$(1\bar{2})$&$(2\bar{1})$&$(2\bar{2})$&$(1\bar{1})$ \\ \hline
$d$&$Y-\frac{1}{2}$&$Y-\frac{1}{2}$&$Y+2L-\frac{3}{2}$&$Y-\frac{3}{2}$ \\ \hline
$c$&$Y-\frac{1}{2}$&$0$&$\frac{1}{2}Y+L-\frac{3}{4}$
&$\frac{1}{2}Y-\frac{3}{4}$ \\
\hline
\end{tabular}
\end{center}
\label{vector}
\caption{\small Vector integrals with 2 denominators}
\end{table}

\begin{table}[H]
\begin{center}
\begin{tabular}{||c|c|c|c|c|c||}
\hline
&$(012)$&$(01\bar{1})$&$(0\bar{1}2)$&$(1\bar{1}2)$&$(01\bar{2})$\\ \hline
$d$&$-2L$&$1-X$&$-L-\frac{1}{4}$&$-1$&$-\frac{1}{2}$\\ \hline
$c$&$-1$&$X-2$&$0$&$-\frac{1}{4}$&$-\frac{1}{2}$\\
\hline
\end{tabular}
\vspace{0.3cm}\\
\begin{tabular}{||c|c|c||c|c||}
\hline
&$(02\bar{2})$&$(12\bar{2})$&$(01\bar{1}2)$&$(012\bar{2})$ \\ \hline
$d$&$2L^2-2L-X$&$1-2L$&$2L-X+1$&$X-2L^2$ \\ \hline
$c$&$2L-2$&$-L+\frac{1}{4}$&$X-1$&$1-2L$ \\
\hline
\end{tabular}
\end{center}
\label{vector2}
\caption{\small Vector integrals with 3 and 4 denominators}
\end{table}

\begin{table}[H]
\begin{center}
\begin{tabular}{||c|c|c|c|c|c||}
\hline
&$(012)$&$(01\bar{1})$&$(0\bar{1}2)$&$(1\bar{1}2)$&$(01\bar{2})$\\ \hline
$g$&$\frac{1}{4}Y+\frac{1}{8}$&$\frac{1}{4}Y-\frac{1}{2}X+\frac{5}{8}$
&$\frac{1}{4}Y$
&$\frac{1}{4}Y-\frac{1}{4}$&$\frac{1}{4}Y+\frac{1}{8}$\\ \hline
$\beta$&$-L$&$-X+\frac{5}{4}$&$-\frac{2}{3}L-\frac{1}{9}$
&$-1$&$-\frac{1}{2}$\\ \hline
$\alpha$&$-\frac{1}{4}$&$-X+\frac{3}{2}$&$0$
&$-\frac{1}{9}$&$-\frac{1}{2}$\\ \hline
$\gamma$&$-\frac{1}{2}$&$2X-\frac{7}{2}$&$0$&$-\frac{1}{4}$&$-\frac{1}{2}$\\
\hline
\end{tabular}
\vspace{0.3cm}\\
\begin{tabular}{||c|c|c||c|c||}
\hline
&$(02\bar{2})$&$(12\bar{2})$&$(01\bar{1}2)$&$(012\bar{2})$ \\ \hline
$g$&$\frac{1}{4}Y+\frac{3}{8}$&$\frac{1}{4}Y$
&$-\frac{1}{2}X+\frac{1}{2}$&$-\frac{1}{4}$ \\ \hline
$\beta$&$2L^2-3L-X+\frac{1}{2}$&$-2L+1$
&$L-X+\frac{5}{4}$&$-2L^2+2L+X-\frac{1}{2}$ \\ \hline
$\alpha$&$L-1$&$-\frac{2}{3}L+\frac{1}{9}$
&$-X+\frac{7}{4}$&$-L+\frac{3}{4}$ \\ \hline
$\gamma$&$2L-\frac{5}{2}$&$-L+\frac{1}{4}$&$2X-3$&$-2L+2$ \\
\hline
\end{tabular}
\end{center}
\label{tensor}
\caption{\small 2-rank tensor integrals with 3 and 4 denominators}
\end{table}

\begin{table}[H]
\begin{center}
\begin{tabular}{||c|c|c|c||}
\hline
&$G^{(1)}$&$G^{(2)}$&$\tau$ \\
\hline
$(01\bar{1}2)$&$\frac{1}{2}X-\frac{7}{8}$&$-\frac{1}{2}X+\frac{5}{8}$
&$\frac{2}{3}L-X+\frac{49}{36}$  \\ \hline
$(012\bar{2})$&$-\frac{1}{8}$&$-\frac{1}{4}$
&$-2L^2+3L+X-1$\\ \hline
 & $\kappa$&$\rho$&$\psi$ \\
\hline
$(01\bar{1}2)$&$X-\frac{29}{18}$&$3X-\frac{19}{4}$
&$-3X+5$ \\ \hline
$(012\bar{2})$&$-\frac{2}{3}L+\frac{11}{18}$
&$-2L+\frac{5}{2}$&$-L+1$ \\
\hline
\end{tabular}
\end{center}
\label{tensor2}
\caption{\small 3-rank tensor integrals with 4 denominators}
\end{table}

\renewcommand{\baselinestretch}{1.0}

\section*{Appendix B. Gauge invariant subset of Feynman diagrams}
\setcounter{equation}{0}
\renewcommand{\theequation}{B.\arabic{equation}}

Amplitudes, describing Feynman diagrams with loop correction to the
real photon emission vertex, and the ones, taking into account self--
energy of fermions (typical diagrams are shown in Fig.1a,b), provide
a gauge invariance in respect to the real photon polarization vector.
It has a universal form and may be taken into account
by substitutions in Born amplitude of the form
\begin{eqnarray}
&& \frac{\hat{p}-\hat{k_1}+M}{-2pk_1}\hat {e} u(p) \to
\frac{\alpha}{2\pi}\left[A_1\left(\hat{e}-
\hat{k_1}\frac{pe}{pk_1}\right)+A_2\hat{k_1}\hat{e}\right]u(p), \nonumber \\
&& A_1=\frac{M}{2(M^2+t)}\left(1-\frac{t}{M^2+t}L_t\right),\qquad t=-2pk_1,
\quad L_t=\log\frac{-t}{M^2}\, ,
\nonumber \\
&& A_2=\frac{N}{t}+\frac{1}{2(t+M^2)}
- \frac{2t^2+3tM^2+2M^4}{2t(M^2+t)^2}L_t\,,\\
\nonumber
&& N=\frac{M^2}{t}\left[\Li(1)-\Li\biggl(\frac{M^2+t}{M^2}\biggr)\right],
\qquad \Li(z)=-\int_0^1\frac{\ln(1-zx)}{x}\dd x,
\end{eqnarray}
where $e$ is the polarization vector of the real photon.
In the IK we have $A_1=1/(4M)$, $\ A_2=(\pi^2/6-1/4)/M^2$.
In a similar fashion for the diagrams, which can be obtained from
depicted in Fig.1a,b by emitting a real photon from another leg, we have
\begin{eqnarray}
\bar{u}(p_2)\frac{\hat{p}_2+\hat{k}_1+m}{2p_2k_1} \to
\frac{\alpha}{2\pi}\bar{u}(p_2)
\left[B_1\left(\hat{e}-\hat{k}_1\frac{p_2e}{p_2k_1}\right)
+B_2\hat{k}_1\hat{e}\right].
\end{eqnarray}
In the IK, omitting the terms disappearing in the limit of zero
electron mass, we have $B_1=0$, $\ B_2=(1/2-2L)/M^2$.

\section*{Appendix C. Averaging on neutrino states, traces}
\setcounter{equation}{0}
\renewcommand{\theequation}{C.\arabic{equation}}

To rearrange bispinors in the matrix element we use Fierz identity:
\begin{eqnarray}
\bar{u}_1O_au_2 \bar{u}_3O_au_4=-\bar{u}_3O_bu_2 \bar{u}_1O_bu_4,\quad
O_a=\gamma_a(1+\gamma_5)/2.
\end{eqnarray}
Summing on the neutrino spin states of the matrix element squared one obtains
\begin{equation}
\Sigma\bar{u_3}O_a u_2(\bar{u_3}O_bu_2)^*=2((q_1q_2)^{ab}-q_1q_2g^{ab})=L_{ab}.
\end{equation}
Averaging over the neutrino momentum is performed using the invariant
integration:
\begin{equation}
\int\frac{\dd^3q_1\dd^3q_2}{q_{10}q_{20}}q_1^aq_2^b\delta^4(q_1+q_2-q)=
\frac{\pi}{6}(q^2g_{ab}+2q^aq^b).
\end{equation}
Application of this formula to the tensor $L^{ab}$ gives the result:
\begin{equation}
\int\frac{\dd^3q_1\dd^3q_2}{q_{10}q_{20}}L^{ab}\delta^4(q_1+q_2-q)=
\frac{4\pi}{3}(q^aq^b-q^2g^{ab})=\frac{4\pi}{3}O^{ab}.
\end{equation}
The doubled interference term of Born and one--loop amplitudes looks
as follows (we consider IK):
\begin{eqnarray}
2\Sigma M_B^*M_1=\frac{2^8\alpha G_F^2\pi}{3M}\biggl[-A_1T_1-A_2T_2+B_2T_3+
\nonumber \\
\frac{1}{2}\int \dd k\biggl[\frac{S_1}{(0)(2)(\bar 1)}+\frac{S_2}{(0)(1)
(\bar 2)}+\frac{S_3}{(0)(1)(2)(\bar 1)}+\frac{S_4}{(0)(1)(2)(\bar 2)}
\biggr]\biggr]\,,
\end{eqnarray}
where the traces are:
\begin{equation}
T_i=\frac{1}{4} {\matr{Tr}}(\hat{T}_i^{ac})O^{ac},\qquad S_i=\frac{1}{4}
{\matr{Tr}}(\hat{S}_{i1}^{ac}- \hat{S}_{i2}^{ac})O^{ac}\,.
\end{equation}
The list of $T_i^{ac}$, $S_{i1}^{ac}$, $S_{i2}^{ac}$ is given below:
\begin{eqnarray*}
\hat{T}_1&=&\hat{p}_2\gamma_a(\gamma_b-2\hat{k}_1p_b/M^2)(\hat{p}+M)
\gamma_b(\hat{p}_2+M)\gamma_{c}/M^4\,,\\
\hat{T}_2&=&\hat{p}+2\gamma_a\hat{k}_1\gamma_b(\hat{p}+M)
\gamma_b(\hat{p}_2+M)\gamma_{c}/M^4\,,\\
\hat{T}_3&=&\hat{p}_2\gamma_a\hat{k}_1\gamma_b\hat{p}
\gamma_{c}\hat{p}\gamma_{c}\hat{p}\gamma_b/M^4\,,
\end{eqnarray*}
\begin{eqnarray*}
S_{31}&=&(\hat{p}+1) \gamma_{c} \hat{p} \gamma_b \hat{p_2} \gamma_{\mu}
(\hat{p}_2-\hat{k}) \gamma_a (\hat{p}_2-\hat{k}+1) \gamma_b (\hat{p}-\hat{k}+1)
\gamma_{\mu}\,,  \\
S_{12}&=&(\hat{p}_2+1) \gamma_b (\hat{p}+1) \gamma_{c} \hat{p} \gamma_b
\hat{p}_2 \gamma_{\mu} (\hat{p}_2-\hat{k}) \gamma_a (\hat{p}_2-\hat{k}+1)
\gamma_{\mu} \,, \\
S_{21}&=&(\hat{p}+1) \gamma_{c} \hat{p} \gamma_b \hat{p}_2 \gamma_b \hat{p}
\gamma_{\mu} (\hat{p}-\hat{k}) \gamma_a (\hat{p}-\hat{k}+1) \gamma_{\mu}\,,  \\
S_{41}&=&(\hat{p}+1) \gamma_{c} \hat{p} \gamma_b \hat{p}_2 \gamma_{\mu}
(\hat{p}_2-\hat{k}) \gamma_b (\hat{p}-\hat{k}) \gamma_a (\hat{p}-\hat{k}+1)
\gamma_{\mu}\,,  \\
S_{32}&=&(\hat{p}+1) \gamma_b (\hat{p}_2+1) \gamma_{c} \hat{p}_2 \gamma_{\mu}
(\hat{p}_2-\hat{k}) \gamma_a (\hat{p}_2-\hat{k}+1) \gamma_b (\hat{p}-\hat{k}+1)
\gamma_{\mu}\,,  \\
S_{42}&=&(\hat{p}+1) \gamma_b (\hat{p}_2+1) \gamma_{c} \hat{p}_2 \gamma_{\mu}
(\hat{p}_2-\hat{k}) \gamma_b (\hat{p}-\hat{k}) \gamma_a (\hat{p}-\hat{k}+1)
\gamma_{\mu}\,,  \\
S_{11}&=&(\hat{p}_2+1) \gamma_b (\hat{p}+1) \gamma_b (\hat{p}_2+1) \gamma_{c}
\hat{p}_2 \gamma_{\mu} (\hat{p}_2-\hat{k}) \gamma_a (\hat{p}_2-\hat{k}+1)
\gamma_{\mu}\,,  \\
S_{22}&=&(\hat{p}+1) \gamma_b (\hat{p}_2+1) \gamma_{c} \hat{p}_2 \gamma_b
\hat{p} \gamma_{\mu} (\hat{p}-\hat{k}) \gamma_a (\hat{p}-\hat{k}+1)
\gamma_{\mu}\,,
\end{eqnarray*}

\begin{eqnarray}
S_1=S_{11}-S_{12},\quad
S_2=S_{21}-S_{22},\quad
S_3=S_{31}-S_{32},\quad
S_4=S_{41}-S_{42}.
\end{eqnarray}

\end{document}